\documentstyle[english,psfig]{cupconf}

\ifoldfss
\else
  \ifnfssone
    \newmathalphabet{\mathit}
      \addtoversion{normal}{\mathit}{cmr}{m}{it}
      \addtoversion{bold}{\mathit}{cmr}{bx}{it}
    \newmathalphabet{\mathcal}
      \addtoversion{normal}{\mathcal}{cmsy}{m}{n}
    \else
    \ifnfsstwo
    \fi
  \fi
\fi

\def\ga{\;\rlap{\lower 2.5pt
 \hbox{$\sim$}}\raise 1.5pt\hbox{$>$}\;}
\def\la{\;\rlap{\lower 2.5pt
   \hbox{$\sim$}}\raise 1.5pt\hbox{$<$}\;}
\newcommand\msun{\rm M_\odot}

\def\hexnumber#1{\ifcase#1 0\or1\or2\or3\or4\or5\or6\or7\or8\or9\or
 A\or B\or C\or D\or E\or F\fi }

\makeatletter
\ifx\CUP@mtlplain@loaded\undefined
\else
\fi
\makeatother
\makeatletter
\ifx\CUP@mtlplain@loaded\undefined
\else
\fi
\makeatother
\def\cm{\rm cm}
\def\K{\rm K}
\def\HH{H$_2$}
\def\tento#1{\times 10^{#1}}
\title[H$_{\it 2}$ Molecules and Structure Formation]{The Role of H$_{\bf 2}$ Molecules
 in Cosmological Structure Formation}

\author[T. Abel \& Z. Haiman]
{T. Abel$^1$ \and \ns Z. Haiman$^2$\footnote{Hubble Fellow}}

\affiliation{$^1$Harvard Smithsonian Center for Astrophysics,
60 Garden Street, Cambridge, MA 02138, USA\\[\affilskip]
$^2$Princeton University Observatory, Princeton, NJ 08544, USA}

\begin{document}
\ifnfssone
\else
  \ifnfsstwo
  \else
    \ifoldfss
      \let\mathcal\cal
      \let\mathrm\rm
      \let\mathsf\sf
    \fi
  \fi
\fi

\maketitle

\begin{abstract}

We review the relevance of ${\rm H_2}$ molecules for structure formation in
cosmology.  Molecules are important at high--redshifts, when the first
collapsed structures appear with typical temperatures of a few hundred Kelvin.
In these chemically pristine clouds, radiative cooling is dominated by ${\rm
H_2}$ molecules. As a result, ${\rm H_2}$ ``astro--chemistry'' is likely to
determine the epoch when the first astrophysical objects appear. We summarize
results of recent three--dimensional simulations. A discussion of the effects
of feedback, and implications for the reionization of the universe is also
given.

\end{abstract}

\firstsection % if your document starts with a section,
\section{Introduction}

\begin{figure} 
% \vspace{6.5cm}
% use \vspace to leave a blank space to glue a figure
% comment \vspace and uncomment next line, il you insert a .ps file
\centerline{\psfig{figure=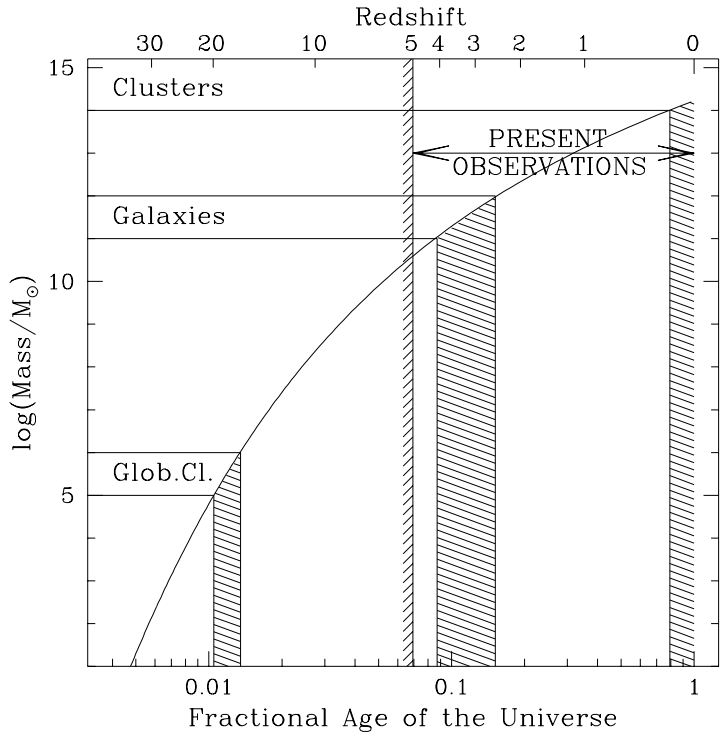,width=10cm,angle=0}}
  \caption{\label{fig:zcollintro} The formation epoch of objects with various
  masses in a standard CDM model ($\Omega=n=\sigma_{8h^{-1}}=1$), shown as a
  fraction of the current age of the universe.  The upper labels indicate
  redshifts. The first objects form at redshift $z\sim 20$, at a ``lookback''
  fraction of $\sim$99\%.}
\end{figure}

In current ``best--fit'' cosmological models, cold dark matter (CDM) dominates
the dynamics of structure formation, and processes the initial density
fluctuation power spectrum $P(K)\propto k^n$ with $n=1$ to predict $n=1$ on
large scales and $n\approx-3$ on small scales (Peebles 1982).  The
r.m.s. density fluctuation $\sigma_M$ then varies inversely with the
mass--scale ($\sigma_M\propto M^{-2/3}$ for $M\gg 10^{12}\msun$, while the
dependence is only logarithmic for $M\ll 10^{12}\msun$).  The more overdense a
region, the earlier it collapses, implying that the present structure was built
from the bottom up, with smaller objects appearing first, and subsequently
merging and/or clustering together to assemble the larger objects (Peebles
1980).  The predicted formation epochs of ``objects'' (i.e. collapsed dark
matter halos) with various masses in the so--called standard CDM cosmology
(Bardeen et al. 1986) are shown in Figure~\ref{fig:zcollintro}.  Galaxies,
which have masses around $10^{11-12}\msun$, are expected to have formed when
the universe had approximately 10\% of its present age (redshift $z\sim3$),
just around the limit of the deepest present--day observations (i.e. the Hubble
Deep Field, HDF, Williams et al. 1996; or Ly$\alpha$ emission line detections,
Weymann et al. 1998).  Clusters of galaxies with masses around
$10^{14-15}\msun$ are predicted to have formed as recently as 80\% of the
current age, with the more massive clusters still assembling at the present
time.

Objects with the masses of globular clusters, $10^{5-6}\msun$, are predicted to
have condensed as early as $\sim$1\% of the current age, or $z\sim 20$.  It is
natural to identify these condensations as the sites where the first
``astrophysical'' objects (stars, or quasars) might be born.  Although the CDM
model in Figure~\ref{fig:zcollintro} predicts still smaller condensations at
even earlier times, the cosmological Jeans mass in the smooth gas after
recombination is $\sim 10^{4}\msun$ (Peebles 1965), implying that gas pressure
inhibits the collapse of gas below this scale (see Haiman, Thoul \& Loeb 1996
on collapse on somewhat smaller scales due to gas/DM shell--crossing, hereafter
HTL96).

What happens in a newly collapsed halo?  Formally, in the absence of
non--gravitational forces, a perfectly spherical top--hat perturbation simply
collapses to a point (Peebles 1980).  According to more accurate treatments
describing self--similar solutions of spherical, but inhomogeneous secondary
infall for a mixture of cold dark matter and baryons (Gunn \& Gott 1972,
Fillmore \& Goldreich 1984, Bertschinger 1985), the evolution is as follows.
In the initial stages, while the bulk of the cloud is still turning around or
expanding, the central, densest region of the cloud collapses and forms a dense
virialized clump.  The dark matter component virializes through violent
relaxation (Lynden--Bell 1967), while the kinetic energy of the gas is
converted into thermal energy through a hydrodynamic shock that raises the gas
temperature to its virial value.  As subsequent gas shells fall towards the
center, they encounter an outward--propagating shock, and are brought to a
sudden a halt. Continuous accretion onto the center then establishes a
stationary, virialized object with a $\rho\propto r^{-2.25}$ density profile.

Early discussions of the formation of galaxies and clusters have argued that
the subsequent behavior of the gas in such a virialized object is determined by
its ability to cool radiatively on a dynamical time (Rees \& Ostriker 1977;
Dekel \& Silk 1986). The same ideas apply on the smaller scales expected for
the first collapsed clouds (Silk 1977; Kashlinsky \& Rees 1983).  Objects that
are unable to cool and radiate away their thermal energy maintain their
pressure support and identity, until they become part of a larger object via
accretion or mergers.  On the other hand, objects that can radiate efficiently
will cool and continue collapsing.

The cooling time is determined by the virial temperature, $T_{\rm vir} \sim
10^4{\rm K}$ $(M/10^8{\rm M_\odot})^{2/3}$ $[(1+z)/11]$.  For the largest
halos, above $T_{\rm vir}\ga10^6$K, the most important cooling mechanism is
Bremsstrahlung; for galaxy--sized halos ($10^4$K$\la T_{\rm vir}\la10^6$K),
cooling is possible via collisional excitation of neutral H and He. As first
pointed out by Saslaw \& Zipoy (1967) and Peebles and Dicke (1968), the virial
temperatures of the first clouds are below $10^4$K. In the primordial gas, the
only molecule that could have a sufficient abundance is ${\rm H_2}$, allowing
cooling between $10^2$K$\la T\la10^4$K (see Dalgarno \& Lepp 1987 for a review
of astrochemistry in the early universe, and Stancil, Lepp and Dalgarno 1996 on
the possible importance of other molecules, such as HD and LiH). Below
$T\la10^2$K, the collapsed clouds are unable to cool, and remain
pressure--supported for longer than a Hubble time.

Although ${\rm H_2}$ molecules are unimportant in the formation of structures
on galactic scales, they likely play a key role in the formation of the first,
smaller structures.  In particular, the abundance of ${\rm H_2}$ controls the
minimum sizes and formation times of the very first systems (HTL96, Tegmark et
al. 1997).

The main issues regarding ${\rm H_2}$ molecules in structure formation,
addressed in this review, are:

\begin{itemize}

\item What is the ${\rm H_2}$ abundance in the first collapsed objects?

\item What is the parameter space (in mass, redshift and metallicity) when
${\rm H_2}$ cooling dominates over all other cooling mechanisms?

\item What is the parameter space when the ${\rm H_2}$ cooling time
is shorter than the dynamical time, so that ${\rm H_2}$ can effect
the dynamics of a system?

\item What feedback mechanisms effect the ${\rm H_2}$ abundance once
the first stars or quasars light up?

\end{itemize}

\section{${\bf H_2}$ Chemistry and Cooling}

Because the cosmological background density of baryons $\sim (\Omega_B
h^2/0.01) (1+z)^3\, 10^{-7} \cm^{-3}$ is very small, chemical reactions in the
smooth background gas occur on long timescales. As a consequence, for dynamical
situations of structure formation chemical equilibrium is rarely an appropriate
assumption.  The dominant \HH\ formation process in the gas phase,
\begin{eqnarray}\label{equ:H2paths}
\rm
H \  \ \ \  + \ \ e^-  \  & \rightarrow & \ \ {\rm H^-} \ \  +  \ \ h\nu,  \\
\rm H^-  \ \ + \ \ H \ \ & \rightarrow & \ \ \rm H_2  \ \ \  +  \ \  e^-,
\end{eqnarray}
relies on the abundance of free electrons to act as catalysts.  At temperatures
low enough to inhibit collisional dissociations by collisions with neutral
hydrogen such electrons can only exist due to non-equilibrium
effects. Electrons are also produced by photoionization of neutral hydrogen
from an external UV radiation field.  It is required to solve the time
dependent chemical reaction network including the dominant chemical reactions.
A very fast numerical method to solve this set of stiff ordinary differential
equations has been developed by Anninos et al. (1997).  The number of possible
chemical reactions involving \HH\ is large, even in the simple case of metal
free primordial gas. The tedious work of selecting the dominant reactions and
their reaction rates has been done by many authors, some recent examples being
HTL96, Abel et al. (1997) and Galli \& Palla (1998, see also Galli and Palla in
this volume).

Cooling (defined here as the radiative loss of internal energy of the gas) is
either due to reaction enthalpy released by a photon, or due to the radiative
decay of collisionally excited atomic or molecular levels. For typical
densities and proto-galactic scales it is accurate to assume the gas to be
optically thin and only consider excitations from atomic and molecular ground
states. The latter fact is due to the low densities resulting in collisional
excitation time scales much longer than the corresponding radiative decay times
(sometimes referred to as the coronal limit). For \HH\ molecules this
assumption breaks down for neutral hydrogen number densities in excess of $\sim
10^{2-3}\cm^{-3}$.  In comparison, for hydrogen atoms the coronal limit is
reached only at electron densities of $\sim 10^{17}\cm^{-3}$ (Abel et al. 1997,
and references therein). The calculations of the appropriate cooling function
for molecular hydrogen seem to be converging. See Flower (2000, this volume)
for a discussion and further references.

\section{${\bf H_2}$ and the First Structures}

Studies that incorporate ${\rm H_2}$ chemistry into cosmological models and
address issues such as non--equilibrium chemistry, dynamics, or radiative
transfer, have appeared only in the past few years.  However, pioneering works
on the effect of ${\rm H_2}$ molecules during the formation of ultra--high
redshift structures go back to the 1960's.  Saslaw \& Zipoy (1967) first
mentioned the importance of ${\rm H_2}$ in cosmology. Peebles \& Dicke (1968)
speculated that globular clusters formed via ${\rm H_2}$ cooling constitute the
first building blocks of subsequent larger structures.  Several papers soon
constructed complete gas--phase reaction networks, and identified the two
possible ways of gas--phase formation of ${\rm H_2}$ via the ${\rm H_2^+}$ or
${\rm H^-}$ channels.  These were applied to derive the ${\rm H_2}$ abundance
in the smooth gas in the post--recombination universe (Lepp \& Shull 1984;
Shapiro, Giroux \& Babul 1994), and under densities and temperatures expected
in collapsing high--redshift objects (Hirasawa 1969; Matsuda et al 1969;
Ruzmaikina 1973).  Palla et al. (1983) combined the molecular chemistry with
simplified dynamics, assuming a uniform sphere in free--fall, finding that
three--body reactions significantly increase the ${\rm H_2}$ abundance at the
later (dense) stages of the collapse.  The significance of non--equilibrium
\HH\ chemistry was realized by Shapiro \& Kang (1987), who studied \HH\
formation in a shock--heated gas, and found that the high electron fraction in
the post--shock region leads to a significantly enhanced \HH\ abundance. Based
on a self--consistent treatment of radiative transfer of the diffuse radiation
field (Kang and Shapiro 1992), this \HH\ enhancement, regulated by
photodissociation inside proto--galaxies, was suggested to lead to the
formation of globular clusters (Kang et al. 1990).

The basic picture that emerged from this papers is as follows. The ${\rm H_2}$
fraction after recombination in the smooth 'intergalactic' gas is small
($x_{\rm H2}=n_{\rm H2}/n_{\rm H}\sim 10^{-6}$). At high redshifts ($z\ga
100$), ${\rm H_2}$ formation is inhibited even in overdense regions because the
required intermediaries ${\rm H_2^+}$ and H$^-$ are dissociated by cosmic
microwave background (CMB) photons.  However, at lower redshifts, when the CMB
energy density drops, a sufficiently large ${\rm H_2}$ abundance builds up
inside collapsed clouds ($x_{\rm H2}\sim 10^{-3}$) at redshifts $z\la 100$ to
cause cooling on a timescales shorter than the dynamical time.  This last
conclusion was found to hold when the rotation of a collapsing sphere was also
included (Hutchins 1976).  Using a different approach, Silk (1983) explicitly
demonstrated that a thermal instability exists for a collapsing gas--cloud
forming ${\rm H_2}$ molecules, leading to fragmentation.  In summary, these
early papers identified the most important reactions for ${\rm H_2}$ chemistry,
and established the key role of ${\rm H_2}$ molecules in cooling the first,
relatively metal--free clouds, and thus in the formation of population III
stars.

The minimum mass of an object that can collapse and cool as a function of
redshift has been studied by Tegmark et al. (1997) assuming constant density
objects, and by HTL96 (see also Bodenheimer and Villere 1986), using
spherically symmetric one--dimensional Lagrangian hydrodynamical models (see
also Bodenheimer and Villere 1986).  Sufficient \HH\ formation and cooling
requires the gas to reach temperatures in excess of a few hundred Kelvin; or
masses of few $\times 10^5~{\rm M_\odot}[(1+z)/11]^{-3/2}$.  Initially linear
density perturbations are followed as they turn around and grow in mass. When
the DM dynamics are included, the center of the collapsing object contracts
adiabatically into the growing DM potential well. As the object grows in mass,
a weak accretion shock is formed. One might imagine a case where the adiabatic
central core forms molecules early and allows rapid collapse before the
accretion shock is formed; however this is not seen in the simulations.  These
temperatures are much larger than the temperature of the intergalactic medium
($T_{IGM}\sim 0.014 (1+z)^2$) at the redshifts of interest. In other words, CDM
models predicts the existence of numerous virialized objects with temperatures
$\la 500\,\K$ that cannot cool. The object that contained the first star in the
universe grows by the infall of both intergalactic DM and of gas, heated in an
accretion shock. The residual fraction of free electrons catalyze the formation
of \HH\ molecules in its central region.

\section{Three-dimensional Numerical Simulations}

Cosmological hydrodynamical simulations of hierarchical models of structure
formation have proven very successful in explaining cosmic structure on
sub-galactic scales (Cen et al. 1994, Zhang et al. 1995, Hernquist et al. 1996,
Dav\'e et al. 1999), galactic scales (e.g. Kauffmann et al. 1999, Katz et
al. 1999) and galaxy clusters (e.g. Frenk et al. 1999, Bryan and Norman 1998).
With a realistic primordial chemistry model (e.g. Abel et al. 1997) and
efficient numerical methods (e.g. Anninos et al. 1997) it is possible to also
simulate the formation of the first cosmological objects.  The first
two--dimensional simulations of structure formation studied the collapse and
fragmentation of cosmological sheets (Anninos and Norman~1996). In these
simulations, the collapsing gas is heated in the accretion shock, and
subsequently found to cool isobarically. The non--equilibrium abundance of
electrons behind the strong accretion shock with $T\gg 10^4\K$ reaches its
maximum value, and fast \HH\ formation up to an abundance of a few $10^{-3}$
(number fraction) is observed, in agreement with the study of \HH\ formation in
shock--heated gas by Shapiro \& Kang (1987).  A similar calculation has been
carried out by Abel et al. (1998b) for the study of the two dimensional
collapse of long cosmic string induced sheets. In this cosmic string scenario,
the dominant mass component was assumed to be hot dark matter (HDM,
e.g. massive neutrinos).  One then envisions long, fast moving cosmic strings
to induce velocity perturbations, causing sheets to collapse, cool and fragment
(Rees 1986). The feedback from the newly formed stars might have then induced
further structure formation as in Ikeuchi and Ostriker (1986). It turns out,
however, that \HH\ formation is inefficient at high redshift where the CMBR
dissociates the intermediaries H$^-$ and H$_2^+$. Furthermore, the shocks
caused by the string--induced velocity perturbations are too weak to enhance
\HH\ formation. Consequently Abel et al. (1998b) concluded that the cosmic
string plus HDM model cannot develop luminous objects before the HDM component
becomes gravitationally unstable.

Numerical calculations of more ``mainstream'' structure formation scenarios
have been presented by Abel (1995), Gnedin and Ostriker (1997, GO97 hereafter),
Abel et al. (1998a, AANZ98), and Abel, Bryan, and Norman (2000, ABN00). The
simulations presented in GO97 focused on simulating the thermal history of the
intergalactic medium and included star formation and feedback mechanisms. Their
simulations were designed to accurately compute the number of the first objects
able to cool and collapse and hence had to sacrifice numerical resolution
within the collapsed objects. Results such as the typical fragment masses,
typical temperatures, etc. are also found in multi--dimensional simulations
that start from more idealized initial conditions (e.g. Bromm, Coppi, and
Larson 1999)

In general cosmological hydrodynamical simulations treat the dynamics
of the assumed collisionless cold dark matter using N--body
techniques. This is typically coupled to a hydrodynamic grid code,
solving the fluid equations for a gas of primordial composition. The
simulations are initialized at a high redshift ($\ga 100$) where
density and velocity perturbations are small and in the linear
regime. The simulation volume typically needs to be chosen rather
small ($\la 1\,$ comoving Mpc) to make sure that the simulation is at
least capable of resolving the baryonic Jeans Mass prior to
reionization,
\begin{eqnarray}
  \label{eq:MJ_IGM}
  M_{J,IGM} \approx 1.0\tento{4}M_\odot\ \left(\frac{1+z}{10}\right)^{3/2} 
  \left(\frac{\Omega_B h^2}{0.02}\right)^{-1/2}. 
\end{eqnarray}
The use of periodic boundary conditions for such small box sizes is only
accurate at relatively high redshifts where they do model a representative
piece of the model universe.  Even in the lowest resolution simulations it
becomes clear that an intricate network of sheets, and dense knots at the
intersection of filaments is found. To the eye this structure is very similar
to the simulation results on much larger scales.

Sufficient \HH\ molecules, enabling the gas to cool, form only in the dense
spherical knots at the intersection of filaments (AANZ98). These knots consist
of virializing dark matter halos that accrete gas mostly from the nearby
filaments but also from the neighboring voids.  An accretion shock transforms
the kinetic energy of the incoming gas into internal thermal energy. This shock
tends to be spherical towards the directions of the voids but is often
disturbed and more complex in morphology at the interface to the filaments. For
the first objects that show any molecular hydrogen cooling the accretion shock
is too weak to raise the ionization level of the gas over its residual
primordial fraction of ${n_{H^+}}/{n_H} \approx 2.4 \times 10^{-4}~
\Omega_0^{1/2} {0.05}/(h\Omega_B)$ Peebles (1993). However, the associated
raise in temperature of the post-shock gas allows molecule formation to proceed
at time scales smaller than the Hubble time.

For the very first (i.e. least massive) objects with virial temperatures $\la
1000\,$K molecule formation is relatively slow, and the cooling time remains
longer than the free fall time. As the objects merge and accrete, the higher
virial temperature allows the chemistry and cooling to operate on faster than
dynamical time scales.  This further merging induces a rather complex velocity
and density field in the gas, as well as the dark matter. Typical cosmological
hydrodynamic methods can not follow the further evolution of the fragmentation
of the gas clouds due to lack of numerical resolution and it was not possible
to asses the nature of the first luminous objects by direct simulation.  This
drawback was recently overcome by the simulations presented in Abel, Bryan and
Norman (2000, ABN00) by exploiting adaptive mesh refinement techniques (Berger
and Collela, 1989, Bryan \& Norman 1997, 1999). This numerical scheme allows to
follow the gas dynamics to smaller and smaller scales by introducing new finer
grids as they are needed.  Since this is done at a scale much below the local
Jeans length one is confident to capture the essential scale of the
fragmentation due to gravitational instability.

These simulations clearly show how a region at the center of the virialized
halo, containing approximately $200\,M_\odot$ in baryons, collapses rapidly.
This ``core'' is formed via the classical Bonnor--Ebert instability of
isothermal spheres. It contracts faster than the dynamical time scales in its
parent halo. Hence these simulations indicate that the first luminous object(s)
(perhaps a massive star) will form before most of the gas in the halo can
fragment (ABN00). This core might still fragment further when it turns fully
molecular via the three body formation process (Palla et al. 1983, Silk
1983)\footnote{Abel, Bryan, and Norman (unpublished) have carried out a
simulation which includes the three--body \HH\ formation covering a dynamic
range of $3\tento{7}$. A preliminary analysis suggests that the core does not
fragment further when it turns full molecular. This suggests that most likely a
massive star will form in the collapsing core.}. If the core forms stars at
100\% efficiency an the ratio of produced UV photons per solar mass is the same
as in present day star clusters than about $6\times 10^{63}$ UV photons would
be liberated during the average life time of massive star ($\sim 5\times
10^{7}\,$yrs). This is a few million times more than the $\sim 10^{57}$
hydrogen molecules within the virial radius, further suppressing \HH\ cooling
and fragmentation. Hence, the first star(s) may halt star formation until the
massive star(s) die(s).  Only a small fraction of primordial gas might be able
to condense into PopIII stars of pristine primordial composition.

\section{Feedback Issues}

The first stars formed via ${\rm H_2}$ cooling are expected to produce UV
radiation, and explode as supernovae (if they are more massive than $\sim 8{\rm
M_\odot}$), producing significant prompt feedback on the ${\rm H_2}$ abundance
in their own parent cloud.  In addition, any soft UV radiation produced below
13.6eV and/or X--rays above $\ga 1$keV from the first sources can propagate
across the smooth H background gas, possibly influencing the chemistry of
distant regions.  Soft UV radiation is expected either from either a star or an
accreting black hole, with a black hole possibly contributing X--rays, as well.
(Although recent studies find that metal--free stars have unusually hard
spectra, these do not extend to $\ga 1$keV. See, e.g. Tumlinson \& Shull 1999).
The most important (and uncertain) quantity for assessing a stellar feedback is
the IMF of the first stars.  Several authors have argued that the IMF might be
(see, e.g. Larson 1999 and references therein) biased towards massive
stars. The lack of zero--metallicity stars, the so--called G--dwarf problem is
resolved if the first generation of stars were short--lived; while the
relatively inefficient cooling of metal--free gas could impose a minimum mass.
A similar conclusion was reached by a recent 3D simulation (Bromm, Coppi, and
Larson 1999).

The key question is whether the ${\rm H_2}$ abundance in a collapsed region is
effected shortly after the first few sources turn on (either in the same
collapsed region, or elsewhere in the universe), i.e. before the ${\rm H_2}$
abundance becomes irrelevant either because objects with $T_{\rm vir}\la 10^4$K
are already collapsing, or because metal enrichment has reached sufficiently
high levels that ${\rm H_2}$--cooling no longer dominates ($\sim 1\%$ solar,
B\"ohringer \& Hensler 1989).  Ferrara (1998) considered the internal feedback
from supernova explosions, and found that the non--equilibrium chemistry in the
shocked gas can increase the ${\rm H_2}$ abundance.  However, Omukai \& Nishi
(1999) argued that a single OB star can photodissociate the ${\rm H_2}$
molecules inside the whole $M\sim10^6{\rm M_\odot}$ cloud. Even if molecules
re-form after SN explosions, they found a net negative feedback.

External feedback from an early soft UV background were considered by Haiman,
Rees \& Loeb (1997).  It was found that ${\rm H_2}$ molecules are fragile, and
easily photo--dissociated even inside large collapsed clumps via the two--body
Solomon process (cf. Field et al. 1966)\footnote{Note that these objects might
also explain the large number of Lyman Limit Systems observed in high redshift
quasar spectra (Abel \& Mo 1998).} -- even when the background flux is several
orders of magnitude smaller than the level $\sim 10^{-21}~{\rm erg\, cm^{-2}\,
s^{-1}\, Hz^{-1}\, sr^{-1}}$ inferred from the proximity effect at $z\sim 3$
(Bajtlik et al. 1988), and needed for cosmological reionization at $z>5$.
These results were confirmed by a more detailed, self--consistent calculation
of the build-up of the background and its effect on the contributing sources
(Haiman, Abel \& Rees 1999).  The implication is a pause in the cosmic
star--formation history: the buildup of the UVB and the epoch of reionization
are delayed until larger halos ($T_{\rm vir}\ga 10^4$K) collapse.  (This is
somewhat similar to the pause caused later on at the 'H-reionization' epoch,
when the Jeans mass is abruptly raised from $\sim 10^{4}~{\rm M_\odot}$ to
$\sim 10^{8-9}~{\rm M_\odot}$.)  An early background extending to the X--ray
regime would change this conclusion, because it catalyzes the formation of
${\rm H_2}$ molecules in dense regions (Haiman, Rees \& Loeb 1996, Haiman, Abel
\& Rees 2000).  If quasars with hard spectra ($\nu F_\nu \approx$ const)
contributed significantly to the early cosmic background radiation then the
feedback might even be positive, and reionization can be caused early on by the
small halos.

\section{Conclusions} 
 
In popular CDM models, the first stars or quasars likely appeared inside
condensed clumps with virial temperatures $T{\rm vir}\la 10^4$K at redshifts
$z\sim 20$.  Because of the low virial temperatures, ${\rm H_2}$ cooling (or
lack thereof) played a dominant role in the gas dynamics inside these
condensations.  State--of--the art numerical simulations identify the sites for
the first star--formation with the intersection of dense filaments (Abel et
al. 1998, 2000).  Although not directly visible in ${\rm H_2}$ emission, these
star--formation sites could be detected out to $z\sim 15$ with the Next
Generation Space Telescope, provided they have a star formation efficiency of
$\ga$ one percent (Haiman \& Loeb 1998), or they form quasar black holes with
an efficiency of $\sim 10^{-3}$ and shine at the Eddington luminosity (Haiman
\& Loeb 1998).  In the latter case, early mini--quasars from ${\rm H_2}$
cooling would reionize intergalactic hydrogen by $z\sim 10$.  The average
star--formation efficiency in collapsed halos before $z\sim 3$ can be estimated
by matching the average metal enrichment of the Ly $\alpha$ forest.  If the
latter is $\sim 10^{-3}$ solar, this implies that $\sim$2\% of the gas mass in
collapsed regions are processed through stars (assuming a ratio in the number
of high--mass to intermediate--mass stars as in a Scalo IMF, see Haiman \& Loeb
1997).  This value is not far from the efficiency of $\sim$1\% suggested by 3D
simulations that resolve sub--parsec scales (ABN00); however it is unlikely
that the early IMF was similar to the one observed in the local universe
(e.g. Larson 1999).  Arguably, further progress towards answering this question
will have to come from a combination of more accurate simulations of
star--formation in a metal--free plasma; including realistic radiative
transfer.

\begin{acknowledgments}

We thank M. Rees for helpful comments. This work was supported by NASA through
a Hubble Fellowship.  TA acknowledges partial support by NSF grant AST-9803137
and NASA grant NAG5-3923.

\end{acknowledgments}

\end{document}